\documentclass[twocolumn]{aastex63}

\submitjournal{ApJ}

\shorttitle{Dust and gas in M\,31 GMCs}
\shortauthors{Forbrich et al.}

\begin{document}

\title{First Resolved Dust Continuum Measurements of Individual Giant Molecular Clouds in the Andromeda Galaxy}

\author{Jan Forbrich}
\affiliation{Centre for Astrophysics Research, University of Hertfordshire,
College Lane, Hatfield AL10 9AB, UK}
\affiliation{Center for Astrophysics $\vert$ Harvard \& Smithsonian, 60 Garden St, MS 72, Cambridge, MA 02138, USA}

\author{Charles J. Lada}
\affiliation{Center for Astrophysics $\vert$ Harvard \& Smithsonian, 60 Garden St, MS 72, Cambridge, MA 02138, USA}

\author{S\'ebastien Viaene}
\affiliation{Sterrenkundig Observatorium, Universiteit Gent, Krijgslaan 281, 9000, Gent, Belgium}
\affiliation{Centre for Astrophysics Research, University of Hertfordshire,
College Lane, Hatfield AL10 9AB, UK}

\author{Glen Petitpas}
\affiliation{Center for Astrophysics $\vert$ Harvard \& Smithsonian, 60 Garden St, MS 72, Cambridge, MA 02138, USA}




\begin{abstract}
In our local Galactic neighborhood, molecular clouds are best studied using a combination of dust measurements, to determine robust masses, sizes and internal structures of the clouds, and molecular-line observations to determine cloud kinematics and chemistry. We present here the first results of a program designed to extend such studies to nearby galaxies beyond the Magellanic Clouds. Utilizing the wideband upgrade of the Submillimeter Array (SMA) at 230~GHz we have obtained the first continuum detections of the thermal dust emission on sub-GMC scales ($\sim$ 15 pc) within the Andromeda galaxy (M~31). These include the first resolved continuum detections of dust emission from individual GMCs beyond the Magellanic Clouds.  Utilizing a powerful capability of the SMA, we simultaneously recorded CO(2-1) emission with identical $(u,\,v)$ coverage, astrometry and calibration, enabling the first measurements of the CO conversion factor, $\alpha_{\rm\,CO(2-1)}$, toward individual GMCs across an external galaxy. Our direct measurement yields an average CO--to--dust mass conversion factor of $\alpha^\prime_{\rm CO-dust} = 0.042\pm0.018$~$M_\odot$ (K km s$^{-1}$ pc$^2$)$^{-1}$ for the $J= 2-1$ transition. This value does not appear to vary with galactocentric radius. Assuming a constant gas-to-dust ratio of 136, the resulting $\alpha_{\rm CO}$ $=$ 5.7 $\pm$ 2.4 $M_\odot$ (K km s$^{-1}$ pc$^2$)$^{-1}$ for the 2-1 transition is in excellent agreement with that of Milky Way GMCs, given the uncertainties. Finally, using the same analysis techniques, we compare our results with observations of the local Orion molecular clouds, placed at the distance of M~31 and simulated to appear as they would if observed by the SMA.
\end{abstract}

\keywords{Andromeda Galaxy -- Giant molecular clouds -- Dust continuum emission -- CO line emission}


\section{Introduction}

Star formation plays a critical role in both galaxy and cosmic evolution. Indeed, it is the rate of star formation that is the primary metric for cosmic evolution studies (e.g., \citealp{mad14}). Within galaxies star formation occurs in giant molecular clouds (GMCs), on scales of several 10~pc. The physical conditions within these clouds set the star formation rate (SFR) for both the clouds and the galaxies which contain them.  Understanding the physical nature of GMCs is thus key to understanding the physical processes that drive star formation, set the SFR and thus determine the evolution of galaxies. 

GMCs are almost entirely composed of molecular hydrogen and  atomic helium gas, mixed with a sprinkling of interstellar dust. These components account for approximately 75\%, 24\% and 1\% of the total cloud mass, respectively in Milky Way GMCs.  Yet, since their discovery nearly five decades ago, almost everything we know about the nature of GMCs had been derived from observations of rare trace molecular species, most notably CO, which account for only about $\sim$ 0.03\% of the total mass of such clouds (e.g., \citealp{hey15}). Measurements of even such fundamental cloud properties as size, mass, and internal structure have been traditionally hampered by uncertainties inherent in such observations, including variations in molecular abundances, opacities and excitation conditions within and between the clouds. 

However, after decades of slow progress in Milky Way molecular cloud research dominated by observations of rare gas traces like CO, the field was transformed when observational capabilities were developed that enabled reliable measurements of the dust in GMCs.  Near-infrared, wide-field imaging surveys produced exquisite dust extinction maps of nearby clouds providing unprecedented data on cloud structure and the first robust and consistently reliable masses of local GMCs (e.g., \citealp{lad94,alv01,goo09,lad10}). The {\it Herschel} and {\it Planck} missions have produced wide-area surveys of dust continuum emission greatly improving on the sensitivity, resolution and dynamic range of dust (and total) column density measurements in Galactic GMCs (e.g., \citealp{and10,pla14,lom14,per16}). 

 Since molecular hydrogen formation occurs on the surface of dust grains (e.g., \citealp{wak17}), dust is expected to be very well mixed with molecular hydrogen gas.
Moreover, measurements of the dust can provide more reliable cloud mass and structural information than those of molecular lines since for GMCs in the Milky Way the dust accounts for $\sim$1\% of total cloud masses ($\sim$ 30 $\times$ that of molecular lines), and measurements of it are not hampered by considerations of excitation, chemistry (e.g., depletion) and opacity that severely hinder molecular-line observations. For example, optically thin dust continuum emission can be well described by a modified blackbody (Planck) function that is given by: $S_\nu = B_\nu (T) \tau_\nu \Omega_\nu$ where $\Omega_\nu$ is the beam solid angle, $\tau_\nu$ the optical depth and $B_\nu(T)$ the usual Planck formula. Determination of a dust column density therefore depends on only two parameters, the dust temperature $T_d$ and the dust opacity $\kappa_\nu$. The cloud mass directly follows from knowledge of the distance to the cloud and an assumed gas-to-dust ratio. 

Over the past decade observations of dust absorption and emission in the Milky Way have greatly improved our understanding of GMCs by providing fundamental new insights into their physical nature and relation to star formation. In particular, such observations have  1)-- provided new clarity in revealing the underlying filamentary nature of GMCs (e.g., \citealp{and10,and14,arz11}), 2)-- shown that molecular clouds are characterized by stratified internal structure (e.g., \citealp{lad99,alv01}), 3) demonstrated that cloud PDFs are very well described by power-law functions \citep{lom15} whose slopes are correlated with the internal level of star formation activity \citep{lad09,kai09}, 4)--described a new, improved,  star formation law relating the star formation rate (SFR) to the amount of dense gas in a cloud \citep{lad10,hei10,eva14}. 5)--dramatically confirmed the \citet{lar81} mass -- size relation for clouds, revealing how exquisitely tight this relation is for local GMCs \citep{lom10}, and 6), showing as a result, that local GMCs cannot be described by a Kennicutt-Schmidt relation of the type that describes star formation on the global scales of galaxies \citep{lad13} and 7)-- produced the first detailed measurements of the cloud core mass functions with shapes similar to the stellar IMF (e.g., \citealp{alv07,kon10}). 

Molecular gas has been observed in external galaxies almost as long as it has been studied in the Milky Way. However, due to the enormous distances involved, these studies have been both dominated by observations of CO (the brightest emitting molecule) and largely restricted to spatial scales (1 - 10 kpc) much larger than those of individual GMCs. With millimeter/submillimeter-wave interferometers, such as ALMA, PdBI and the SMA,  resolved measurements of GMCs are possible in the closest galaxies and CO studies of GMC populations in these galaxies now approach being placed on a footing similar to that of Milky Way studies (e.g., M 51, \citealp{col14}; NGC 300, \citealp{fae18}).
 
Similar to the situation in the Milky Way, measurements of resolved dust emission from extragalactic GMCs would result in significant advances in our understanding of the nature of these objects.  Dust continuum observations on the scales of 100-200 pc in the nearest spiral galaxies, such as M31 and NGC 300 have been enabled by the {\it Herschel} and {\it Planck} space missions (e.g., \citealp{fri12,rie18}), but resolved dust continuum observations of GMCs  on scales of $\sim$10~pc have remained elusive until now. Such spatial resolution in the millimeter wavelength range can be achieved using existing interferometers, but so far, it has been impossible to reach the required combination of continuum sensitivity, angular resolution and spatial filtering to detect and resolve dust emission from individual GMCs beyond the Magellanic Clouds. 

The recent upgrade of the continuum receivers on the Submillimeter Array (SMA) is now providing the first opportunity to obtain resolved measurements of dust emission from individual GMCs in another galaxy: the Andromeda Galaxy, M\,31. At a distance of  780~kpc, M31 is the nearest large disk galaxy to the Sun.  With a bandwidth of currently 32~GHz at an observing frequency of 230~GHz \citep{gri16}, the SMA has the required continuum sensitivity to detect dust emission from GMCs in M31. Moreover, its $(u,v)$ coverage is such that these GMCs can be perfectly matched to a synthesized beam size of $\sim$10~pc and a maximum angular scale of $\sim100$~pc. These capabilities put us in a unique position to detect and resolve, for the first time, the dust emission in individual GMCs from an external galaxy beyond the Magellanic Clouds. Coincidentally, it is in M~31 that the first resolved CO measurements of extragalactic GMCs were originally obtained \citep{lad88,vog87}.

Moreover, the CO(2-1) emission line can be extracted from the wide continuum band to both remove CO contamination of the expected weak continuum signal from the dust and simultaneously produce a discrete observation of the CO gas emission {\it with identical $(u,v)$ coverage, spatial scales, and calibration as the dust observations} along with relatively high spectral (velocity) resolution. While it would be possible to select the observing band such that continuum sensitivity is maximized by excluding CO, recording both has several advantages, in spite of the challenges due to fact that the dust continuum emission is very faint in comparison. Most importantly, due to the higher relative sensitivity, the CO data can be used to delineate the clouds and determine their structure, which in turn helps to identify any contaminant continuum emission that is not related to clouds.

Given the fortuitous match of the spatial resolution, spatial filtering, and sensitivity of the SMA to relevant absolute scales at the distance of M~31, and the latitude of the SMA coupled with  M31's Northern declination, it would be hard to replicate these capabilities at any other existing observatory. In particular, since ALMA cannot observe M~31, the SMA observations we report here open a new and unique window into our most massive, neighboring galaxy.

The Andromeda galaxy is comparable in overall mass, size and metallicity to the Milky Way but is curiously different in a number of interesting ways. Unlike the Milky Way, M31 has weak spiral structure and its star formation is largely confined to a well-defined ring of molecular gas and dust with a galactocentric radius of $\sim$ 10 kpc, the center of which is offset from the galaxy's center (e.g., \citealp{dam93,gor06}). This ring is suspected to be the product of a head on collision between M31 and a smaller galaxy (M32) that took place 210 million years ago \citep{blo06}. The global star formation rate (SFR) is estimated to be between 0.25 - 0.7 M$_\odot$ yr$^{-1}$ for M31, a factor of $\sim$ 2-6 lower than that estimated for the Milky Way (i.e., 1.65 M$_\odot$ yr$^{-1}$; \citealp{lic15}). This firmly places M31 below the so-called  main sequence of star forming galaxies (e.g., \citealp{pen10}). The total molecular mass of M31 is $\sim$ 3 $\times$ 10$^8$ M$_\odot$ \citep{dam93,nie06} and is less than the total molecular mass of the Milky Way (i.e., 10$^9$ M$_\odot$; \citealp{hey15}) by about a factor of three. The most comprehensive studies of M31's star-forming clouds in molecular gas \citep{nie06} and dust \citep{kir15} probe scales of $\sim$ 90 pc, sufficient to resolve GMC complexes but not individual GMCs. Resolved measurements of both gas and dust emission from individual GMCs in M31 should lead to new and more detailed tests of our fundamental understanding of both the nature of GMCs and the laws of star formation in galactic disks.

In the following, we describe observations obtained with the SMA, from what is now an ongoing large-scale science program to survey the dust continuum emission of a significant population of GMCs in M31. We describe the first sub-GMC scale detections of dust continuum emission from individual GMCs in M31. We also report the first spatially resolved measurements of dust continuum emission from individual GMCs in an external galaxy beyond the Magellanic clouds. In this first analysis and paper, we focus on GMC-scale measurements of the CO conversion factor, $\alpha_{\rm CO}$, across the galaxy by linking the continuum and CO measurements. To put these data in context, we describe simulated observations of what the nearby Orion GMCs would look like if we placed them at the distance of M\,31 and observed them with the upgraded SMA.

\section{Source Selection, Observations, calibration, and data reduction}
\label{sec_obs}

Our source selection was based on the results of the {\it Herschel} HELGA dust continuum survey. In particular, we selected our target list from the catalog of Giant Molecular Associations (GMAs) in M31 \citep{kir15}. We initially selected 100 candidate regions based on the observed Herschel fluxes and a range of galactocentric radii ( $\sim$ 6 - 16 kpc) of the GMAs to insure we sampled potentially strong sources over a representative range of environmental conditions. Over the course of the early survey we found that the likelihood of detecting dust continuum emission on $\sim$ 15 pc scales was not strongly correlated with the strength of the Herschel emission obtained on $\sim$ 100 pc scales, nor was it particularly sensitive to galactocentric position.

The observations reported here were carried out with the Submillimeter Array (SMA) 
in its subcompact configuration during 
two campaigns from UT 2018 Aug 16 to UT 2018 Sep 08 then
again during the period of UT 2019 Sep 09 to UT 2019 Oct 16. Since we
expected the CO emission to be strong, we optimized our tuning to
minimize the amount of the band contaminated by CO. Observations were
performed using RxA and RxB receivers tuned to LO frequencies of
225.55 GHz and 233.55 GHz respectively, which placed $^{12}$CO $J$=2-1
in spectral window S1 of RxA USB and $^{13}$CO and C$^{18}$O $J$=2-1 in spectral window S1 of RxA LSB,
leaving RxB free from CO. This configuration provides 32 GHz of
continuous bandwidth ranging from 213.55 GHz to 245.55 GHz with a
spectral resolution of 140.0 kHz per channel, with the CO
contamination contained within only S1 of RxA. For quick looks at the
data, RxA S1 was flagged and completely removed from the band to derive 28 GHz of CO free
continuum measurements, but the final images presented here were
created using all 32 GHz of the correlator with just the individual
$^{12}$CO and $^{13}$CO-contaminated {\it channels} of RxA S1 flagged and removed.

Simple experiments comparing in-band fluxes before and after removal of CO emission show that CO emission typically accounts for 10 - 50\% of the total observed flux in the full continuum band, emphasizing the critical importance of removing CO emission in order to obtain reliable measurements of broad-band (dust) continuum emission.

For all observations the nearby quasars 0013+408 and 0136+478 were
used as the primary phase and amplitude gain calibrators with absolute
flux calibration performed primarily by comparison to Neptune at the
start of the observation, and occasionally cross checked using Uranus
at the end of the observation should the early data be of poorer
quality. Passband calibration was derived using 3c454.3. The raw data
were smoothed to 1.5~km\,s$^{-1}$ and calibration was performed using the MIR
IDL package for the SMA, with subsequent analysis performed in MIRIAD.

To smooth out any variations in sensitivity caused by weather,
generally three targets were observed during a single track (with
typically 6.3 hours shared on the targets), with that observation
being repeated three times. Using this technique we were able to
achieve a relatively uniform rms across all sources of 0.25 mJy or
better. For the observing period in 2018, the SMA used all 8 antennas
in subcompact configuration resulting in an average synthesized beam of
$4.5''\times3.8''$ (or $\sim15$~pc), whereas during the 2019 campaign, one of the two "outer
ring" antennas was in the hangar for repairs resulting in a
synthesized beam more on the order of $8''\times5''$ (or 30$\times$19~pc).
The $^{12}$CO(2-1) and $^{13}$CO(2-1) transitions were excised from the wideband data, and imaged separately. 
For this study we produced integrated intensity, moment~0, maps of the excised $^{12}$CO(2-1) data, and we defer a full discussion of the CO data to a follow-up paper.

\begin{deluxetable*}{llrrrrr}
\tablenum{1}
\tablecaption{Cloud sample and properties \label{tbl_sample}}
\tablewidth{0pt}
\tablehead{
\colhead{Cloud ID} & \colhead{$F_{\rm cont}$} & \colhead{rms} & \colhead{$I_{\rm CO}$   } & \colhead{$M_{\rm dust}$} & \colhead{$\alpha^\prime_{\rm CO-dust}$} & \colhead{$R_{\rm gal}$}  \\
\colhead{        } & \colhead{(mJy)         } & \colhead{(mJy)} & \colhead{(K km s$^{-1}$)} & \colhead{($M_\odot$)   } & \colhead{$M_\odot$ (K km s$^{-1}$ pc$^2$)$^{-1}$} & \colhead{(kpc)}
}
\decimalcolnumbers
\startdata
K026A & 1.86$\pm$0.31 &  0.23 & 20.4$\pm$1.1 &  529$\pm$ 89 & $0.035\pm0.008$ &  5.8 \\ 
K026B & 1.17$\pm$0.23 &  0.23 &  6.0$\pm$0.8 &  333$\pm$ 65 & $0.150\pm0.040$ &  5.8 \\ 
K048  & 0.86$\pm$0.27 &  0.27 & 18.0$\pm$1.1 &  244$\pm$ 77 & $0.040\pm0.013$ &  5.7 \\ 
K092  & 1.43$\pm$0.25 &  0.21 & 11.5$\pm$0.9 &  406$\pm$ 71 & $0.027\pm0.007$ &  8.0 \\ 
K093A & 1.13$\pm$0.20 &  0.20 & 11.0$\pm$0.7 &  322$\pm$ 58 & $0.084\pm0.021$ &  8.1 \\ 
K093B & 0.64$\pm$0.20 &  0.20 & 13.5$\pm$0.7 &  182$\pm$ 58 & $0.039\pm0.012$ &  8.1 \\ 
K134  & 1.27$\pm$0.31 &  0.31 &  7.3$\pm$1.8 &  362$\pm$ 88 & $0.051\pm0.021$ & 10.5 \\ 
K136  & 2.17$\pm$0.36 &  0.25 & 10.5$\pm$1.6 &  617$\pm$101 & $0.031\pm0.010$ & 11.3 \\ 
K157  & 1.70$\pm$0.31 &  0.31 & 12.6$\pm$1.3 &  483$\pm$ 89 & $0.039\pm0.009$ & 11.9 \\ 
K160A & 0.99$\pm$0.22 &  0.22 &  5.5$\pm$0.9 &  282$\pm$ 62 & $0.053\pm0.015$ & 12.3 \\ 
K160B & 0.70$\pm$0.22 &  0.22 &  7.5$\pm$0.9 &  200$\pm$ 62 & $0.027\pm0.009$ & 12.3 \\ 
K162A & 1.20$\pm$0.22 &  0.19 & 27.7$\pm$0.8 &  342$\pm$ 62 & $0.017\pm0.004$ & 11.8 \\ 
K162B & 0.64$\pm$0.19 &  0.19 & 14.9$\pm$0.7 &  181$\pm$ 53 & $0.024\pm0.007$ & 11.8 \\ 
K170A & 1.47$\pm$0.29 &  0.23 & 27.0$\pm$1.5 &  417$\pm$ 84 & $0.024\pm0.006$ & 11.8 \\ 
K170B & 1.00$\pm$0.23 &  0.23 & 17.6$\pm$1.1 &  285$\pm$ 65 & $0.043\pm0.011$ & 11.8 \\ 
K176  & 1.71$\pm$0.29 &  0.24 & 11.3$\pm$0.6 &  486$\pm$ 84 & $0.030\pm0.007$ & 13.8 \\ 
K190  & 1.63$\pm$0.22 &  0.24 &  4.6$\pm$0.2 &  463$\pm$ 63 & $0.122\pm0.020$ & 13.9 \\ 
K191  & 6.41$\pm$0.71 &  0.17 & 34.5$\pm$0.7 & 1826$\pm$202 & $0.053\pm0.007$ & 12.1 \\ 
K213A & 3.91$\pm$0.52 &  0.31 & 22.2$\pm$0.7 & 1115$\pm$147 & $0.049\pm0.008$ & 12.0 \\ 
K213B & 3.37$\pm$0.46 &  0.31 & 14.2$\pm$0.6 &  959$\pm$131 & $0.085\pm0.015$ & 12.0 \\ 
K213C & 1.03$\pm$0.31 &  0.31 & 16.9$\pm$0.4 &  294$\pm$ 89 & $0.048\pm0.015$ & 12.0 \\ 
K291A & 1.09$\pm$0.22 &  0.22 &  5.1$\pm$0.2 &  312$\pm$ 62 & $0.162\pm0.036$ & 16.0 \\ 
K291B & 0.69$\pm$0.22 &  0.22 &  1.3$\pm$0.2 &  196$\pm$ 62 & $0.410\pm0.132$ & 16.0 \\ 
\enddata
\end{deluxetable*}

\begin{figure*} 
 
\begin{minipage}{0.32\linewidth} 
\includegraphics[width=\linewidth]{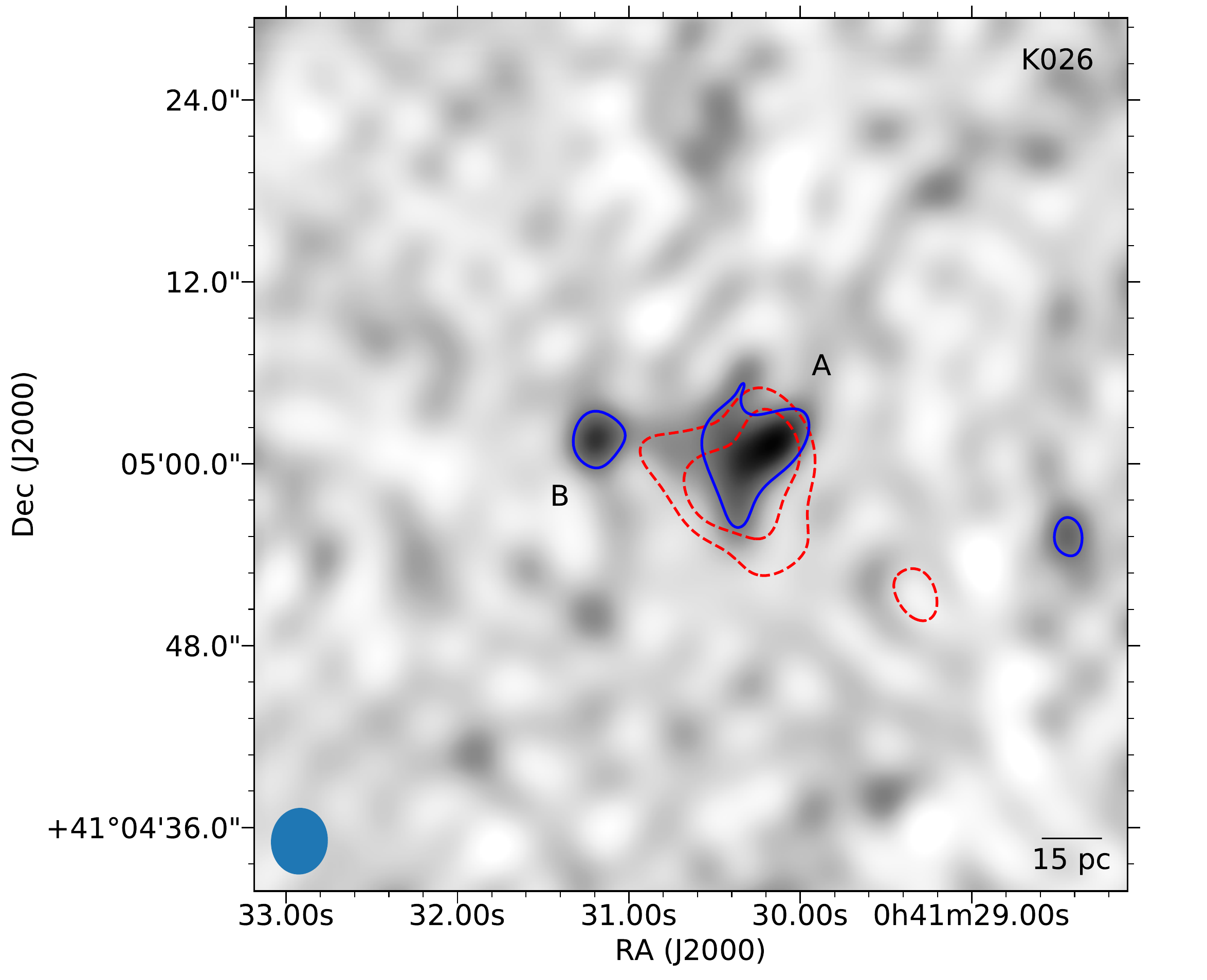}
\end{minipage}  
\begin{minipage}{0.32\linewidth} 
\includegraphics[width=\linewidth]{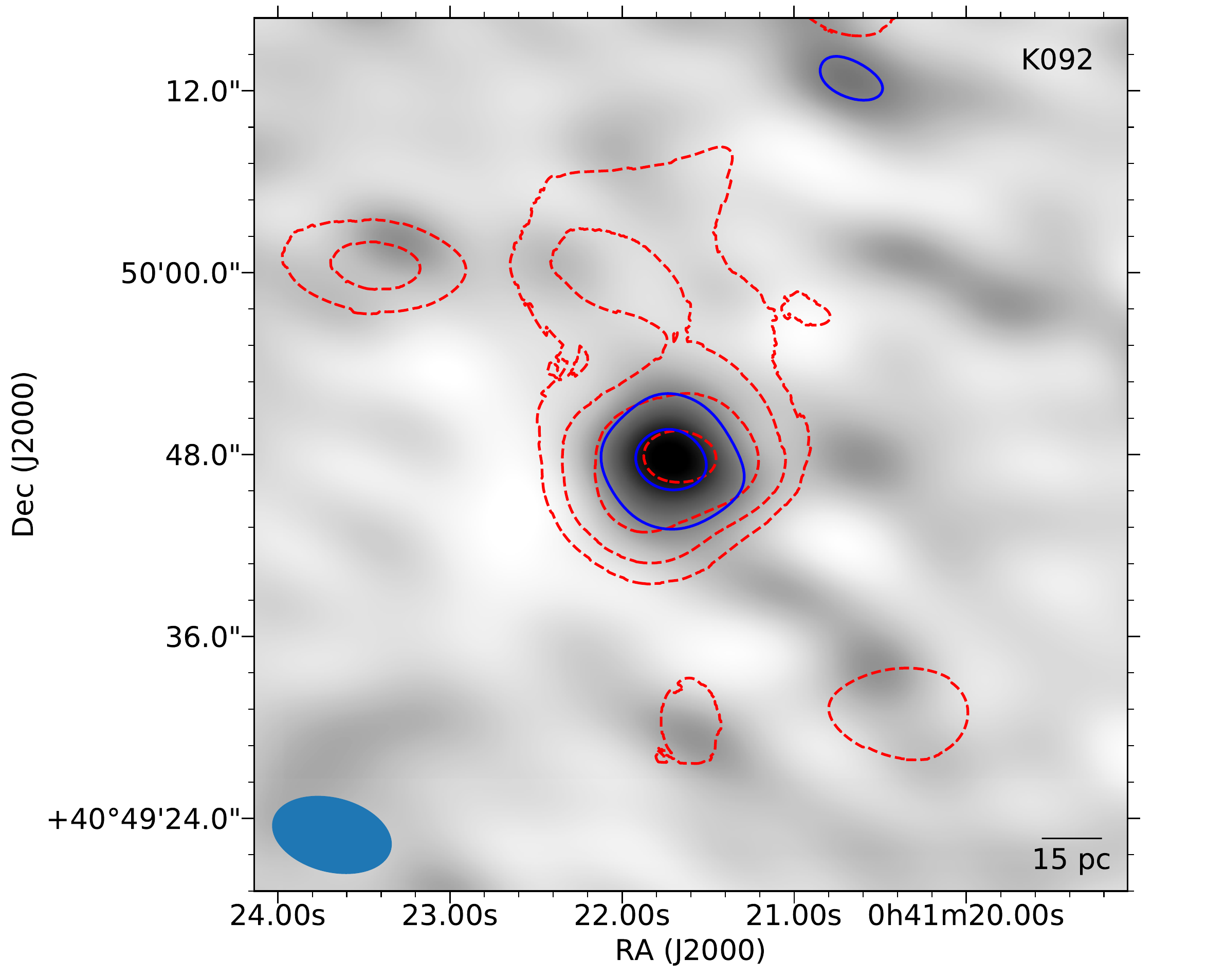}
\end{minipage}  
\begin{minipage}{0.32\linewidth} 
\includegraphics[width=\linewidth]{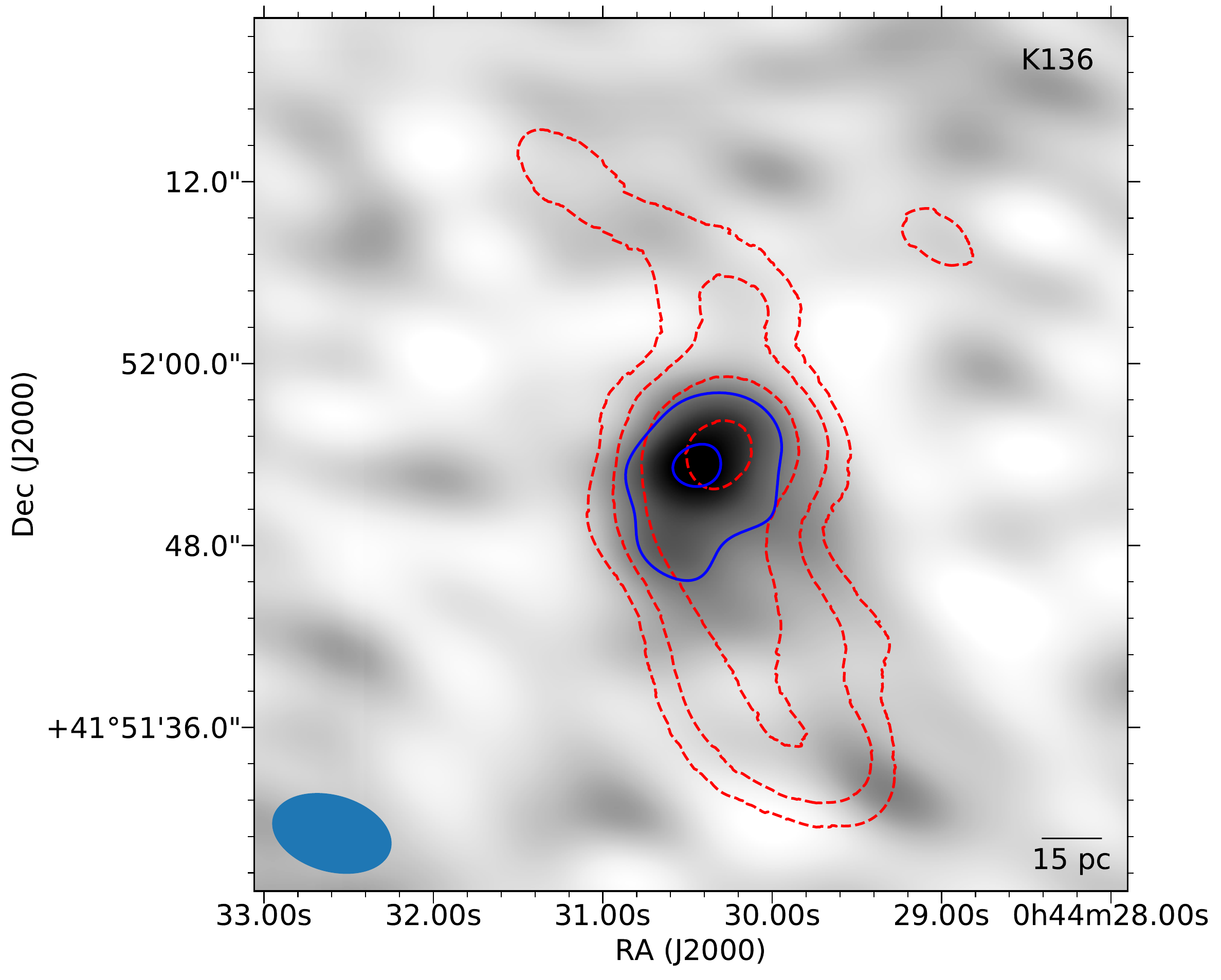}
\end{minipage}  

\begin{minipage}{0.32\linewidth} 
\includegraphics[width=\linewidth]{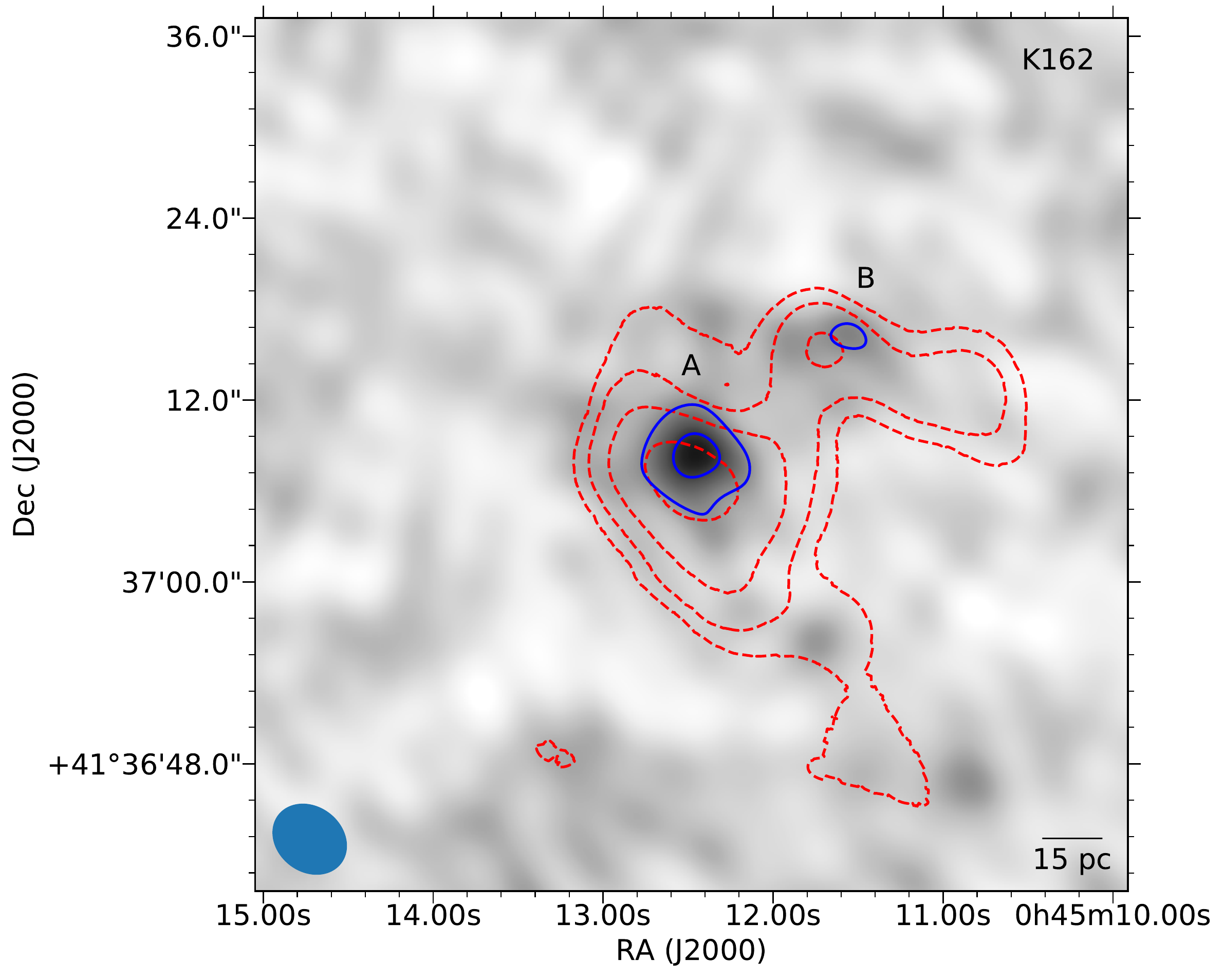}
\end{minipage}  
\begin{minipage}{0.32\linewidth} 
\includegraphics[width=\linewidth]{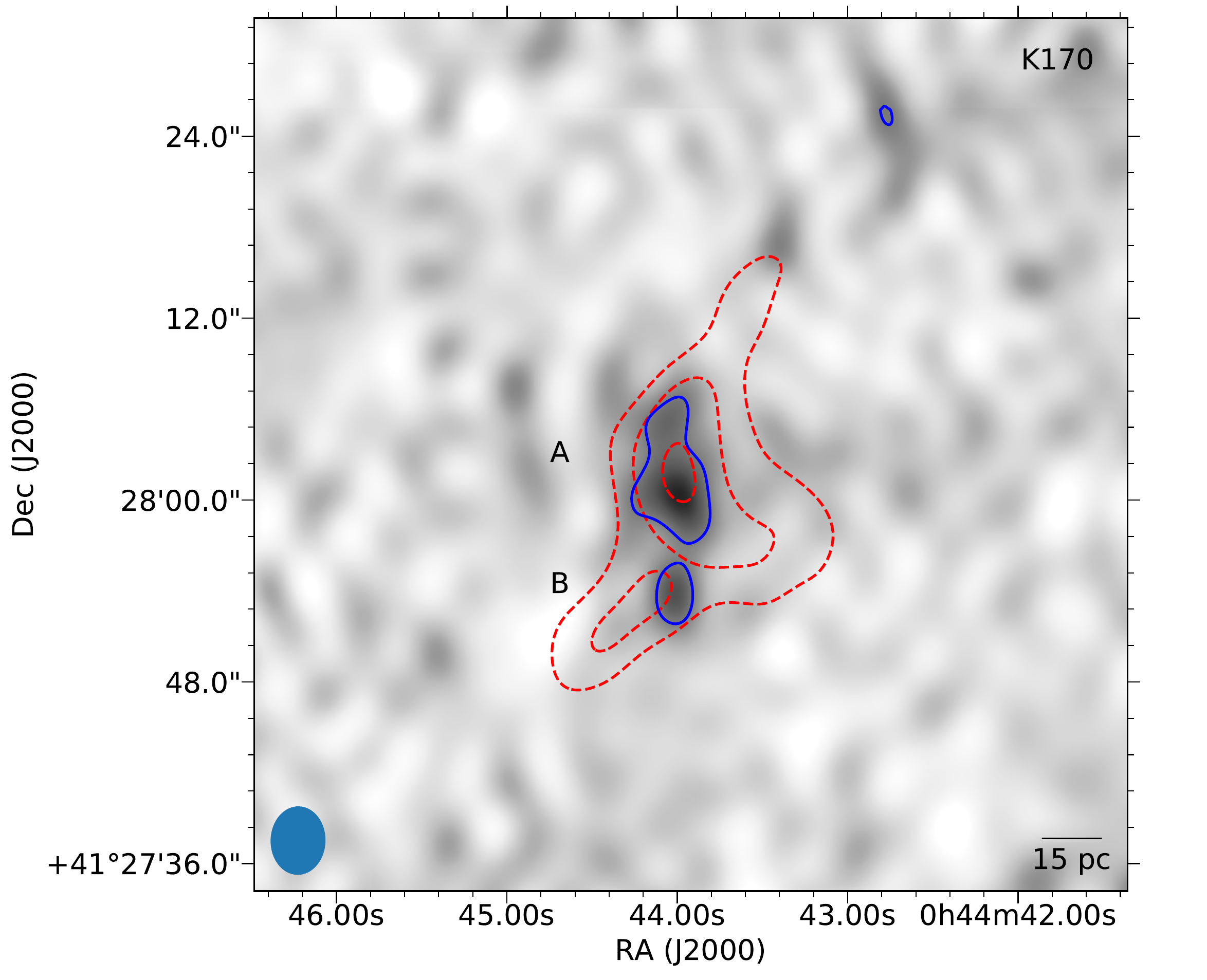}
\end{minipage}  
\begin{minipage}{0.32\linewidth} 
\includegraphics[width=\linewidth]{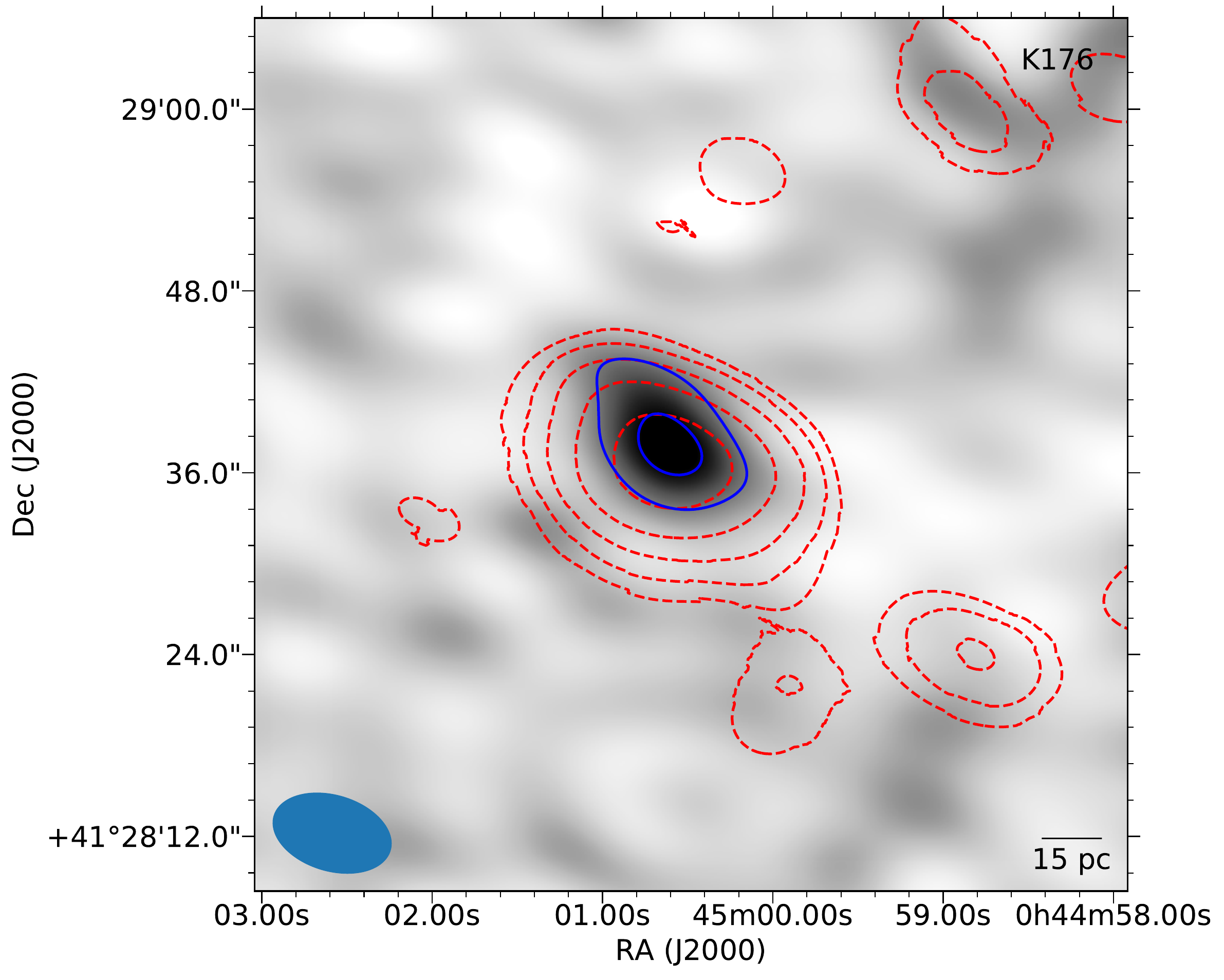}
\end{minipage}  

\begin{minipage}{0.32\linewidth} 
\includegraphics[width=\linewidth]{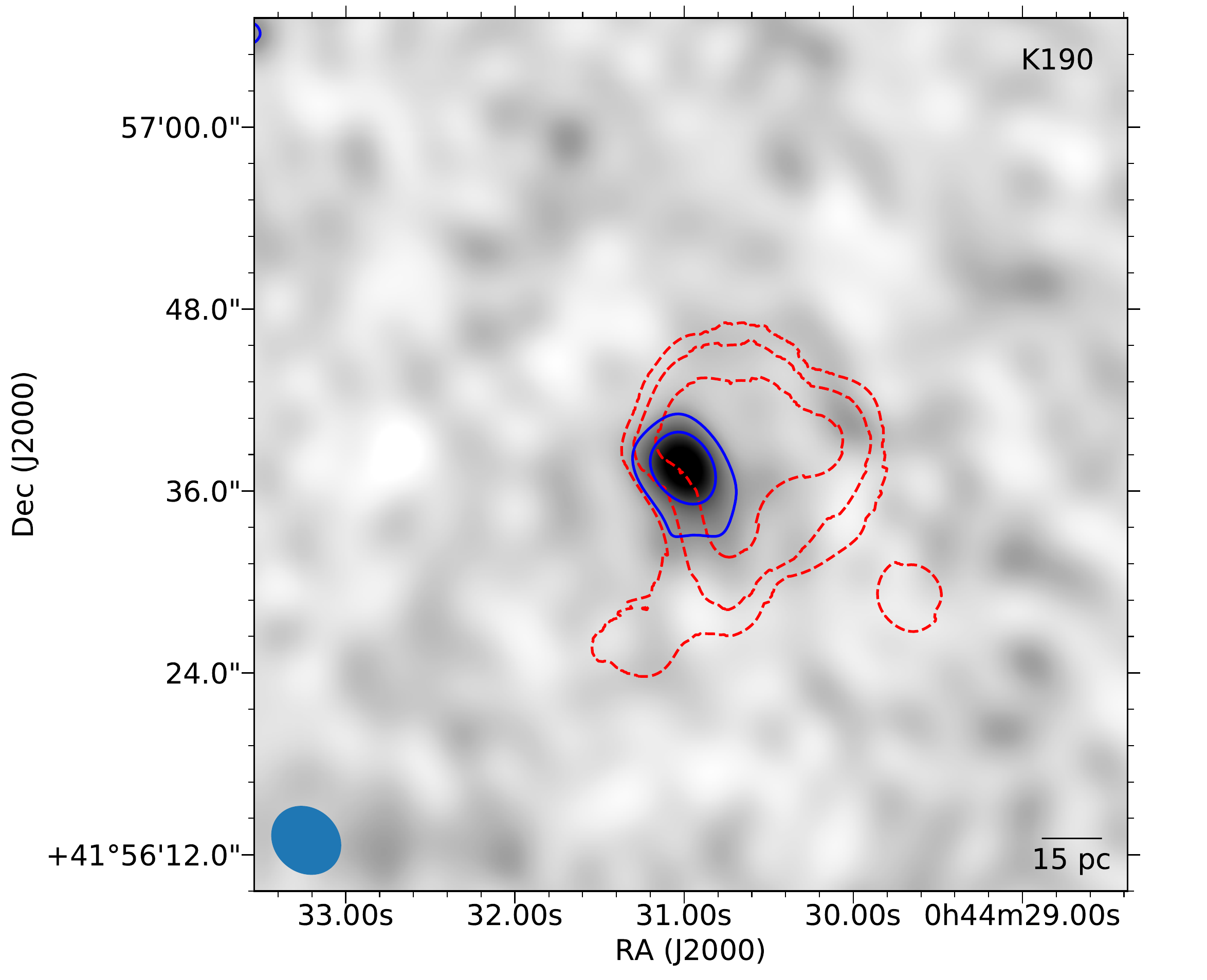}
\end{minipage}  
\begin{minipage}{0.32\linewidth} 
\includegraphics[width=\linewidth]{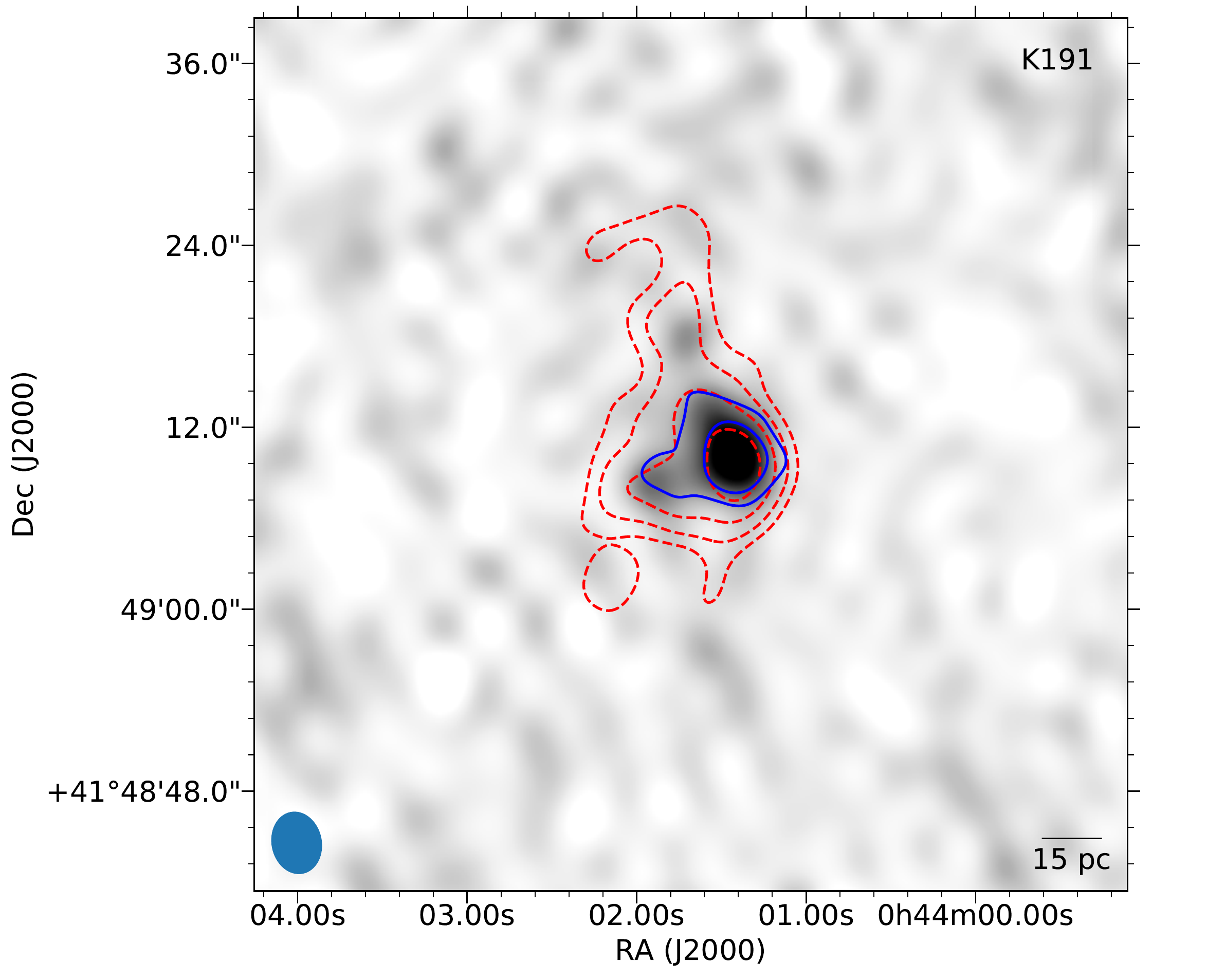}
\end{minipage}  
\begin{minipage}{0.32\linewidth} 
\includegraphics[width=\linewidth]{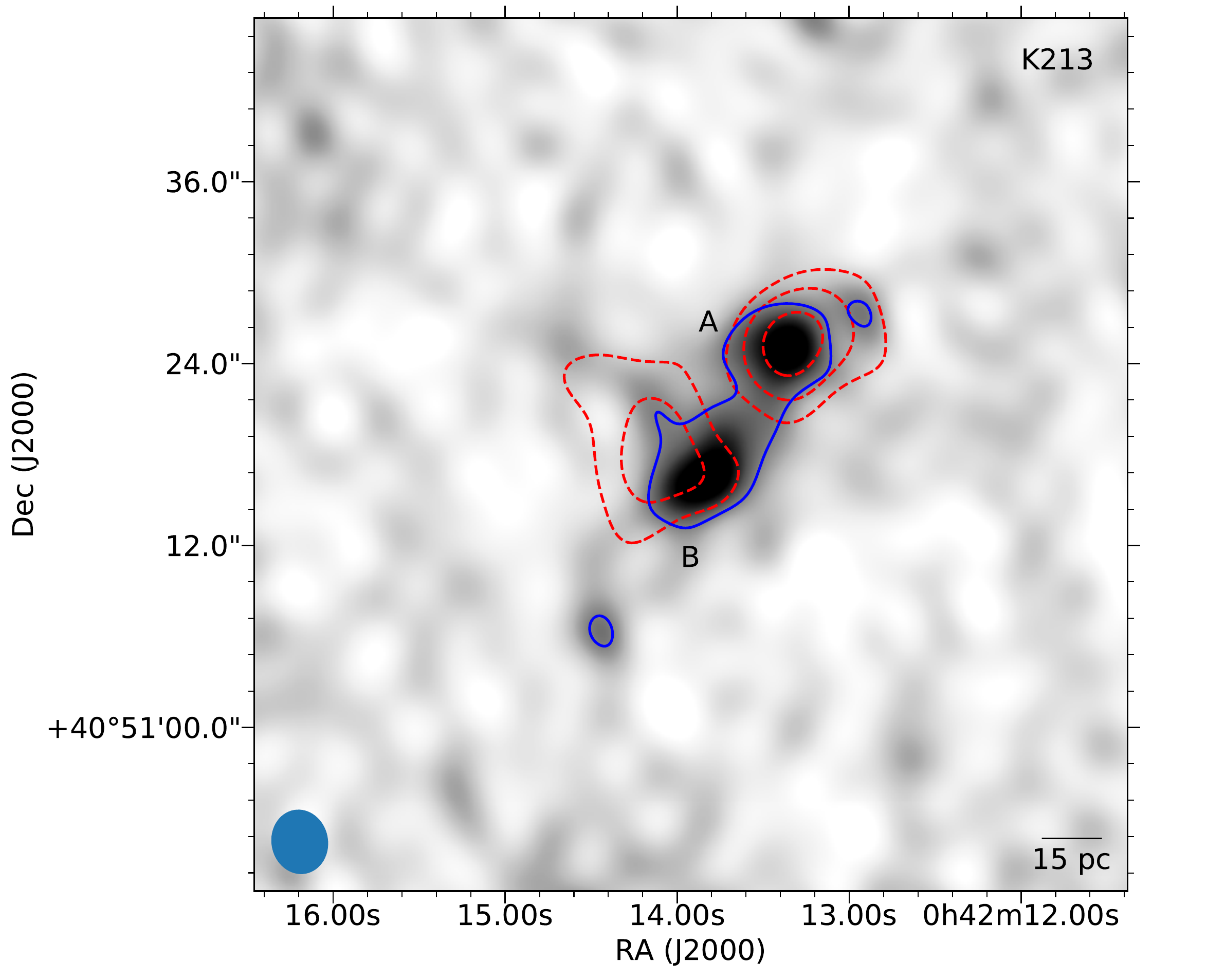}
\end{minipage} 

\caption{Continuum images (greyscale) of individual GMCs with resolved dust continuum emission, at identical scale (typical resolution of 15~pc shown by scalebar), indicated by the blue contour lines (3$\sigma$). For orientation, the red dashed contour lines additionally indicate the $^{12}$CO moment~0 map in steps of 6, 12, 24, etc. $\sigma$. The synthesized beam size is indicated in the lower left corner of each panel.\label{fig_contCO}}

\end{figure*}

\section{Data Analysis and Results}

In this paper our analysis will focus on comparing the dust continuum detections with corresponding CO emission to measure the CO conversion factor, $\alpha_{\rm CO}$, for the cloud sample thus far observed in our program. 

For this purpose, we have analyzed images of the continuum data after removing the small CO emitting portions of the continuum band. We look for continuum emission above a significance limit of 3$\sigma$, where the sensitivity achieved in the individual observations varies slightly (see Section~\ref{sec_obs}). In our initial sample, all but one of our targets (K161) show significant dust continuum emission. The ability to separately excise potentially contaminating CO(2-1) emission from the wideband SMA data allows us to use both continuum and separate CO images to identify and characterize the GMCs. Since the relative sensitivity is far greater in CO than in the dust continuum, the CO moment~0 maps represent the best resolved information on the location, size, and shape of the GMCs in the sample, resolving the {\it Herschel} GMA candidates into one or several clouds. The CO and continuum maps also have identical astrometry, and the CO maps thus also generally set the outer bounds for continuum detections within the confines of a GMC. As an example, the resulting continuum images for nine targets with resolved continuum detections are shown in Figure~\ref{fig_contCO} along with the corresponding contours of CO emission. Given the unusual fractional bandwidth of these observations, we consider a source to be resolved if the emitting area of significant dust emission is larger than the synthesized beam by a factor of 1.2.  In most cases the CO emission is more extended than the dust emission as expected given the current disparity in relative sensitivities to cloud material of the two tracers. The coincidence or near-coincidence of peak continuum emission with peak CO emission confirms that we are observing emission from dust at these positions in the clouds.

To facilitate a direct comparison of dust and CO emission, we use the regions with significant continuum emission to define masks for the extraction of the corresponding CO data. In this way, we extract corresponding continuum and CO fluxes for all targets. We express the CO fluxes in units of $I_{CO}$ (i.e., K km s$^{-1}$). Many of the sources have several spatially distinct detections of significant dust continuum emission and once identified we assess these individual detections separately, labeling the resulting individual sources in each target image with letters (A, B, C, etc.). Out of a total of 15 initial targets included here, we thus end up with a total of 23 continuum sources. For each of these, we obtain measurements of the integrated continuum and $^{12}$CO(2-1) emission, and these are listed in Table~\ref{tbl_sample}. For resolved sources, these two measurements are integrated in the same mask, while otherwise unresolved per-beam measurements at identical positions are compared. The continuum flux density for every cloud can be converted into a dust mass, i.e., 

\begin{equation}
    M_{dust} = S_\nu d^2/(\kappa_\nu B_\nu(T_d))
\end{equation}

To evaluate equation 1 we need to first specify the dust opacity, $\kappa_\nu$, and this requires a choice of a dust model. We adopt the canonical THEMIS Milky Way dust model \citep{jon17} which was calibrated using the {\it Planck} data and other observational constraints by \citet{ysa15}. At 230 GHz, this becomes $\kappa_\nu = 0.0425$ m$^2$kg$^{-1}$. The THEMIS dust model proved to work well in the radiative transfer model of M31 by \citet{via17}, which links dust extinction to dust emission. However, the kind of dust model most suitable for M31 is still a matter of debate (see, e.g., \citealp{whi19}).

Assuming a dust temperature of $T_d=20$~K and a distance of 780 kpc, the dust masses for the clouds calculated from equation 1 with the above assumptions range from 181 to 1825 $M_\odot$. To obtain total cloud masses requires knowledge of the gas-to-dust ratio, $R_{g2d}$, that is, $M_{tot}=  M_{dust} \times R_{g2d}$. Assuming a dust-to-gas ratio of $\sim$136 to compare with the Milky Way would result in a range of total cloud masses of about 24,000 to 240,000 $M_\odot$. For a different assumed gas-to-dust ratio the total mass would be $M_{tot}' = M_{tot} \times [R_{g2d}/136]$.

\section{Discussion}
 
 As described above, observations of dust in both absorption and emission provide the best tool for measuring many of the most fundamental properties of GMCs and star formation. However, in many situations such as in the distant universe and even the more remote regions of the Milky Way, CO will remain the best available tracer of GMCs and the molecular ISM for the foreseeable future. To be able to make full use of its capabilities as a tracer of molecular gas it is essential to be able to calibrate its effectiveness by direct comparison with observations of dust across as wide a range of conditions as possible.  

\subsection{Measuring $\alpha_{\rm CO}$ in M31 GMCs}

Observations of CO are most frequently used to derive molecular gas and cloud masses. In particular, it is assumed that the luminosity of CO from a region of space is directly proportional to the mass of molecular gas in that region. This is often expressed as: $M_{\rm tot} = \alpha_{\rm CO} \times L_{\rm CO}$. Here $M_{tot}$ is the total molecular mass, including a 36\% correction for the presence of heavier elements (i.e., helium, etc.) and has the units of solar masses (M$_\odot$), $\alpha_{\rm CO}$ is the CO conversion factor and has units of $M_\odot$ (K km s$^{-1}$ pc$^2$)$^{-1}$ and  $L_{CO}$ is the CO luminosity expressed in the units of K km s$^{-1}$ pc$^2$. The value of $\alpha_{\rm CO}$ for the 1-0 transition of CO has been both theoretically (using virial arguments) and empirically (often from direct comparison with dust) derived, and it is estimated to have a value of 4.3 $M_\odot$ (K km s$^{-1}$ pc$^2$)$^{-1}$  in the Milky Way and throughout the local universe \citep{bol13}. This corresponds to a value of 6.1 $M_\odot$ (K km s$^{-1}$ pc$^2$)$^{-1}$ for the $J = 2-1$ transition of CO, assuming a typical intensity ratio of I$_{\rm CO}$(2-1)/I$_{\rm CO}$(1-0) of 0.7. Variations in $\alpha_{\rm CO}$ have been reported in the nuclear regions of the Milky Way and other galaxies and are expected in regions of varying metallicity (\citealp{bol13} and references therein). 

Our observations of dust and CO in M 31 provide an opportunity to measure $\alpha_{\rm CO}$ on sub-GMC scales across the Andromeda galaxy. Measuring the value of $\alpha_{\rm CO}$ is straightforward from our data, given the identical spatial sensitivity, astrometry, and calibration of the dust continuum and CO data.

\begin{figure*} 
\includegraphics[width=\linewidth]{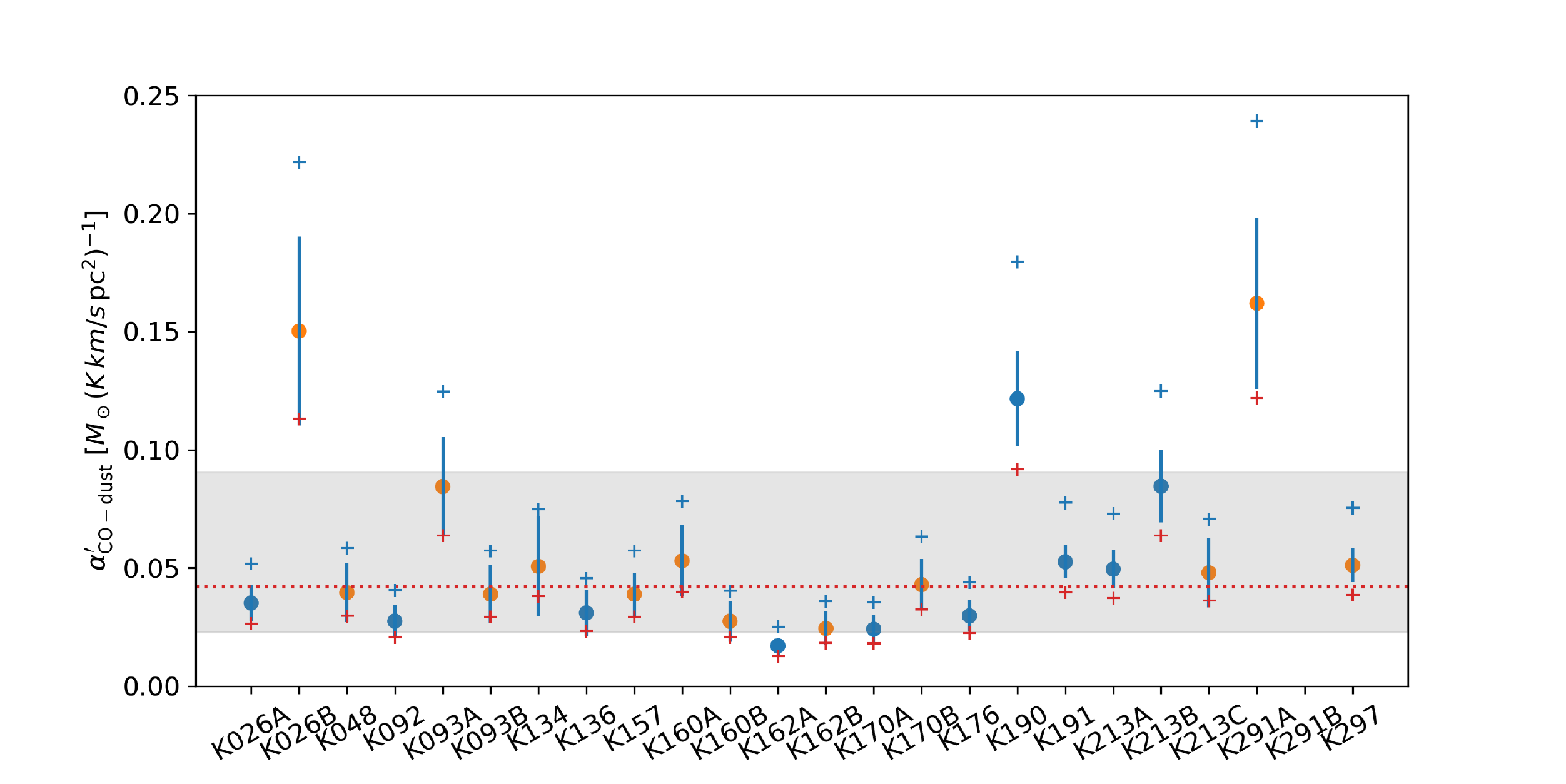}
\caption{Direct $\alpha^\prime_{\rm CO-dust}$ measurements per cloud, converting CO luminosity to dust mass. Blue dots indicate resolved sources, and orange dots indicate unresolved sources, both for $T_d=20$~K. Red and blue crosses additionally show $\alpha_{\rm CO}$ for $T_d=25$~K and $T_d=15$~K, respectively. The dotted line indicates the (clipped, see text) sample mean of $\alpha^\prime_{\rm CO-dust}=0.042$~$M_\odot$ (K\,km\,s$^{-1}$pc$^{-2}$)$^{-1}$. For comparison, we divide the nominal Galactic value for the CO(2-1) transition (see text) by a gas-to-dust ratio of 136 to find a near-identical value of $\alpha^\prime_{\rm CO-dust (gal)} = 0.045$~$M_\odot$ (K\,km\,s$^{-1}$pc$^{-2}$)$^{-1}$, and we indicate the corresponding range with an uncertainty of $\pm0.3$~dex as a gray-shaded area. Note that K291B is off-scale (see Table~\ref{tbl_sample}). \label{fig_alphaPrimeCO}}
\end{figure*}

Our most direct measurement is that of a conversion factor from CO luminosity to dust mass, $\alpha^\prime_{\rm CO-dust}$, since it does not depend on the choice of a gas-to-dust ratio. We first obtain CO measurements that correspond to the mask where significant dust continuum emission has been detected, with the initial hypothesis that all of these objects may be GMCs. By calculating the CO luminosity in this same mask, we can then directly measure the $\alpha^\prime_{\rm CO-dust}$ parameter. In the case of multiple dust continuum peaks per target, we consider each of these separately. The results are shown in Table~\ref{tbl_sample} and displayed in Figure~\ref{fig_alphaPrimeCO}. In spite of a considerable range in galactocentric radius already in this preliminary sample, we find that the clouds can be described by a single value of $\alpha^\prime_{\rm CO-dust}$. Excluding the four most extreme outliers, which are likely contaminants (see below), we find a mean value of $\alpha^\prime_{\rm CO-dust} = 0.042\pm0.018$~$M_\odot$ (K km s$^{-1}$ pc$^2$)$^{-1}$.
Assuming a gas-to-dust ratio of $\sim$136 to compare with Milky Way clouds, this value corresponds to $\alpha_{\rm CO (M31)} = 5.7 \pm 2.4$~$M_\odot$~ (K km s$^{-1}$ pc$^{2}$)$^{-1}$. 
Within the uncertainties our result is in good agreement
 with the Galactic value of $\alpha_{\rm CO (gal)} = 6.1$~$M_\odot$ (K\,km\,s$^{-1}$pc$^{2}$)$^{-1}$ for the 2-1 transition of CO \citep{bol13}.
Our result is also compatible with the study of \citet{ler11} whose estimates of $\alpha_{CO}(1-0)$ on kpc scales in M31 were also reported to be consistent with the Milky Way value. 

These considerations depend on the assumptions spelled out above, most notably the assumed dust temperature (see Equation 1). While all our target GMAs have dust temperatures determined by the HELGA survey, as listed in \citet{kir15}, these estimates are limited by the angular resolution of the {\it Herschel} SPIRE-350$\mu$m band at $\sim24''$ or $\sim90$~pc, smoothing out a lot of the temperature structure present at resolutions of $\sim15$~pc. We have thus decided to assume a common typical dust temperature of $T_d=20$~K. However, the impact of different temperatures on the determination of $\alpha_{\rm CO}$ is limited. This is shown in Figure~\ref{fig_alphaPrimeCO}, where we additionally plot the $\alpha^\prime_{\rm CO-dust}$ values for assumed temperatures of $T_d=15$~K and $T_d=25$~K. 

Nonetheless, given our assumptions, our measurements are compatible with a constant $\alpha^\prime_{\rm CO}$ value for the  M\,31 GMC population we have so far sampled. This value is in excellent agreement with that (i.e., 6.1/136 $=$ 0.045) of Milky Way GMCs and is quite interesting given the considerable range in galactocentric radii (6 to 12 kpc) and presumably environments spanned by our target GMCs in M31. 
Similarly, our derived estimate for $\alpha_{\rm CO}$ also is compatible with a constant value, provided the gas-to-dust ratio is constant in the regions we have observed. 

Finally, we note that there are also a few outliers with seemingly very high $\alpha^\prime_{\rm CO}$. These are sources where unresolved continuum emission is detected in areas of only very faint or no CO emission, in spite of the high sensitivity of the CO observations. As we discuss below these objects are real and likely contaminants of differing physical nature from the sources of dust emission within the GMCs.

\subsection{Nature of the Contaminating Sources of Continuum Emission}

The common astrometry of both datasets enables us to spatially match continuum peaks to structure in the CO maps with a high degree of confidence that would be difficult to duplicate in experiments where the two datasets were obtained at different epochs or with different telescopes. This is a great advantage because it not only allows accurate association of dust and CO emission, but also enables confident identification of any continuum sources not spatially related to the GMCs traced by CO.  For example, in a few of our images we find continuum peaks that are offset from the locations of the CO emitting gas. The much higher relative sensitivity of the CO data to cloud material suggests that such sources are not associated with GMCs in M31 and are real contaminants of a different physical nature or class than the dust sources we detect within the GMCs. Without the angular resolution and common astrometry provided by the SMA, such continuum emission could be falsely associated with the CO emission.

Potential sources of `contamination' include the direct detection of free-free emission from luminous H\,{\sc ii} regions, even at 230~GHz, and the detection of background galaxies. While the latter can only be corroborated if suitable additional data are available, we can at least approximately assess the incidence of H\,{\sc ii} regions whose free-free emission is  bright enough to be detected at 230~GHz. As a first estimate, we can scale the centimeter-wavelength free-free flux density of the Orion Nebula to the distance of M\,31, applying a spectral index of $-0.1$ for optically thin free-free emission to extrapolate to 230~GHz. The total flux density of this relatively puny H\,{\sc ii} region of $\sim400$~Jy (e.g., \citealp{fel93}) corresponds to an unresolved source of 0.09~mJy at the distance of M\,31, and it is thus not detectable in our observations. However, a source with just seven times the flux density of the Orion Nebula would be marginally detectable at 3$\sigma$, assuming an rms sensitivity of 0.20~mJy. Next we now place the Orion Nebula into the Galactic context. While no unbiased catalog of H\,{\sc ii} regions exists for either the Milky Way or M\,31, we can use the catalog obtained by \citet{mur10} in an all-sky survey for free-free emission from H\,{\sc ii} regions based on data obtained by the Wilkinson Microwave Anisotropy Probe (WMAP). They list the free-free luminosities of 183 Galactic H\,{\sc ii} regions. Even though necessarily incomplete, this list contains 174 regions that are more luminous than the Orion Nebula, and 149 regions that are at least seven times and up to three orders of magnitude more luminous. While the full Galactic context remains unclear, the detection, in our observations, of free-free emission from H\,{\sc ii} regions at 230~GHz and in M\,31 thus is evidently possible.

The four outliers in our $\alpha_{\rm CO}$ results, namely K026B, K190, and K291A/B, are candidates for non-dust continuum emission. Out of these, K026B has very little CO emission, and it coincides with a faint X-ray \citep{wan16} and a faint, red {\it Spitzer} source \citep{kha17}. It thus seems plausible that K026B is in fact a background galaxy. The other three sources will require further analysis to establish their nature.

\begin{figure*} 
 
\begin{minipage}{0.5\linewidth} 
\includegraphics[width=\linewidth]{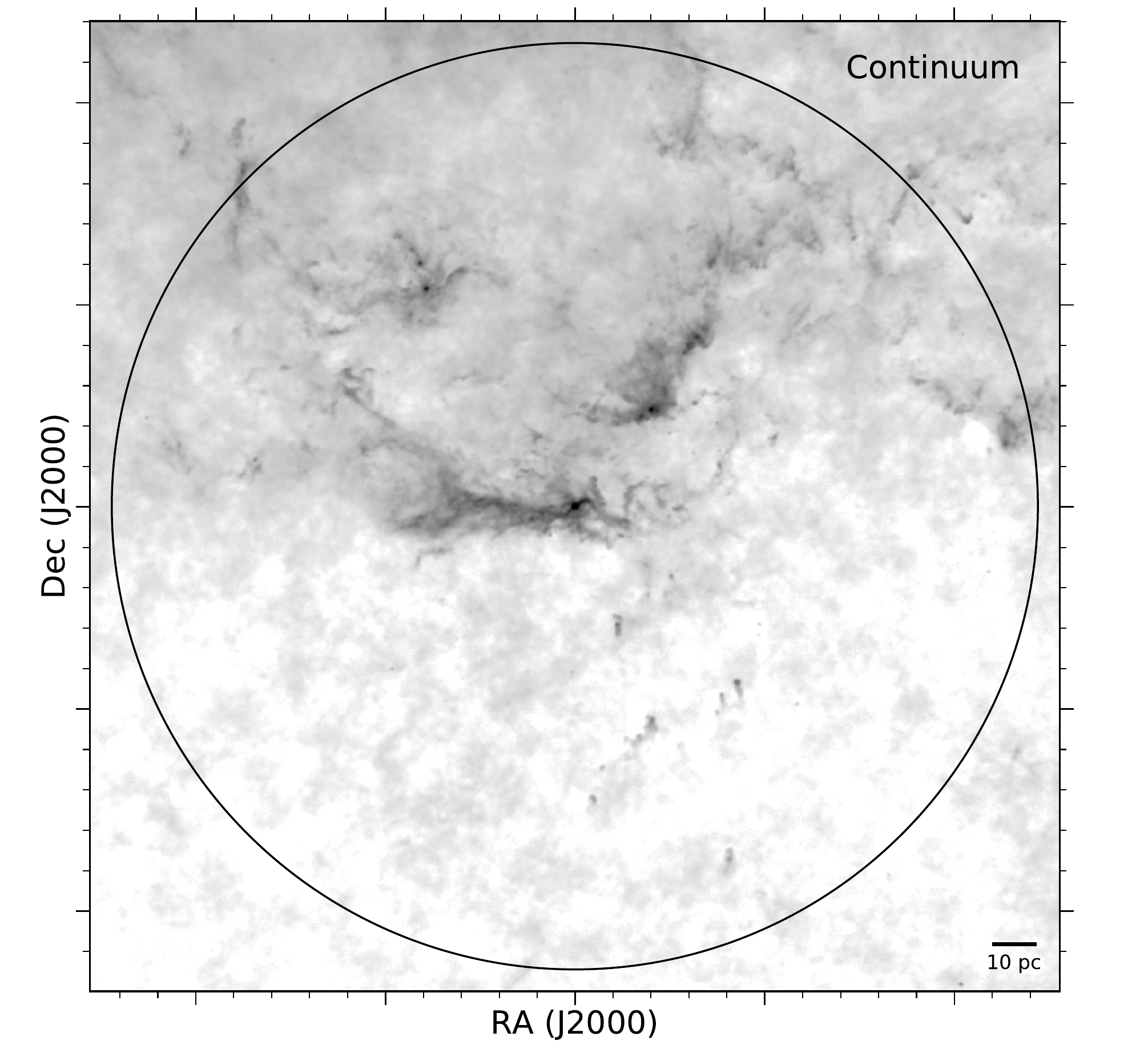}
\end{minipage}  
\begin{minipage}{0.5\linewidth} 
\includegraphics[width=\linewidth]{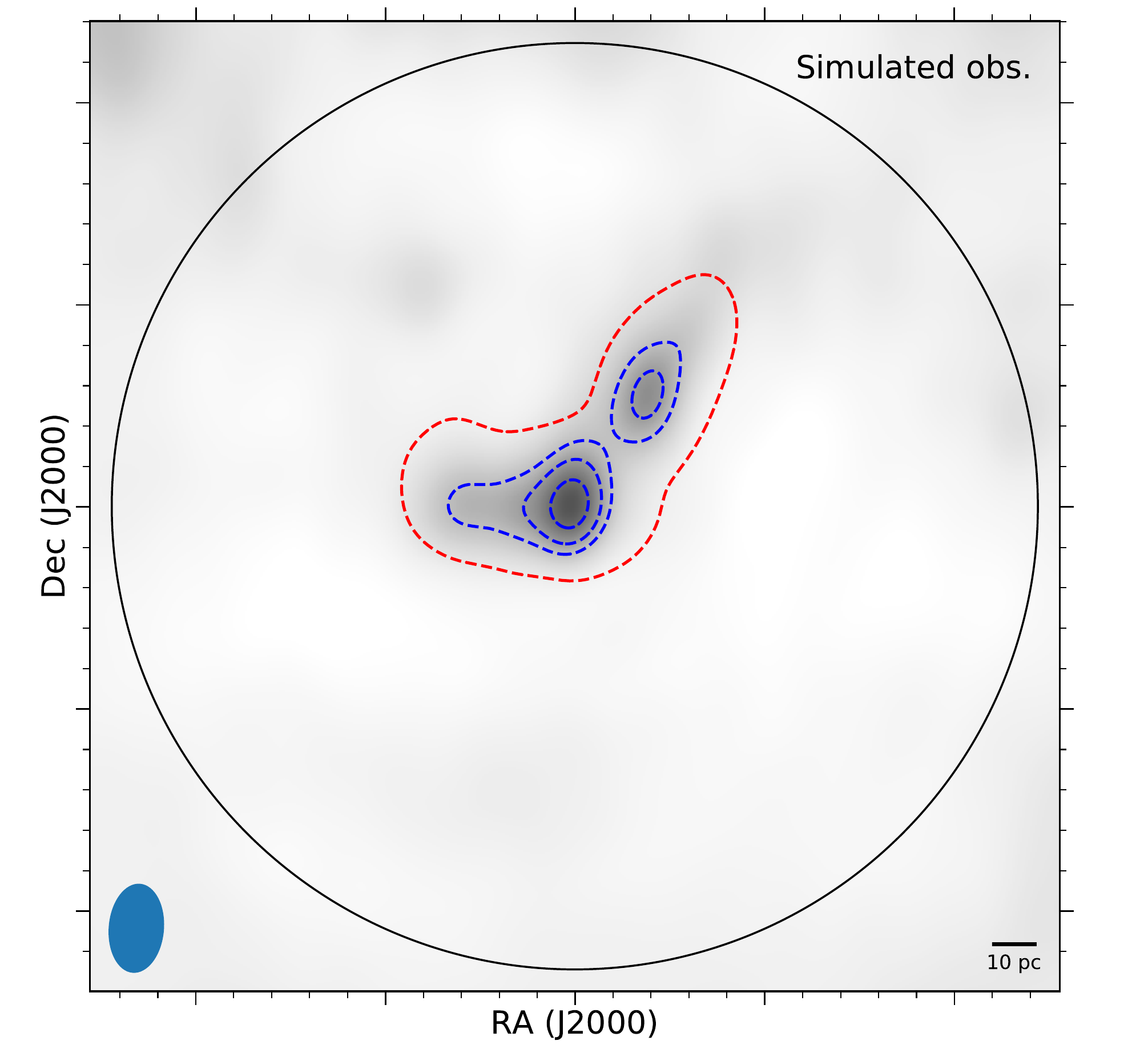}
\end{minipage}  

\caption{Simulated 230~GHz SMA observations of the Orion molecular clouds, {\it as if they were located in M\,31}, assuming the subcompact configuration, with the HPBW primary beam indicated (see text). Left: input {\it Planck} 217~GHz continuum image. Right: simulated continuum observation, with blue contours indicating a typical 3$\sigma$ per-track sensitivity with 1$\sigma$=0.20~mJy\,bm$^{-1}$, for 1, 2, and 4 tracks with a bandwidth of 32~GHz. To visualize these different observing setups in the same image, no artificial noise has been added, such that the simulated image retains the full dynamic range of the model image. The red contour indicates the S/N=100 contour for a single track of the simulated CO(2-1) observations. \label{fig_simori}}
\end{figure*}

\subsection{Context: Orion GMCs at the Distance of M31}

To assess the types of clouds that we are detecting in M\,31, it is helpful to put the results into the context of the local clouds. How would the Orion molecular clouds appear in our observations, if they were located in M\,31? To answer this question, we use existing dust continuum and CO maps of the Orion clouds. 
The {\it Planck} satellite has provided us with all-sky maps that are ideal for this purpose. For a simulation that is as close as possible to our actual observations, we opt for the map at 217~GHz. These are bolometric maps with a much wider effective bandwidth than the SMA, but they are ideal for a general test of feasibility in terms of both S/N and spatial recovery of the cloud. In a first step, we convert the units of the {\it Planck} map from K(CMB) to MJy\,sr$^{-1}$. Given the wide {\it Planck} band, different spectral energy distributions (SEDs) can make a difference in this conversion, either in the form of a correction factor, or as a change in the effective frequency. In the case of a typical dust SED, the correction is accomplished by a shift in effective frequency to 229~GHz, which is perfect for our purposes. Alternatively, at the same reference frequency, the correction factor is only 0.85 \citep{pla14a}. The map is then converted into Jy\,beam$^{-1}$ before we convert the pixel scale, astrometry, and beam size of the map to reflect the distance of M\,31 rather than Orion (414~pc; \citealp{men07}). A direct comparison in CO(2-1) is provided by \citet{nis15}, who mapped the Orion clouds in various molecular transitions with a 1.85~m dish \footnote{See also \url{http://www.astro.s.osakafu-u.ac.jp/~nishimura/Orion/}}.

For our purposes of simulating observations of Orion-like clouds in M\,31, both datasets have far better angular resolution than required and are thus ideal. To simulate observations, we first converted the {\it Planck} data from K(CMB) to Jy\,bm$^{-1}$ by applying a conversion factor of 483.690~MJy\,sr$^{-1}$ \citep{pla14a} at an effective frequency of 229~GHz (see above), and assuming a beam size of 5$'$ (FWHM). Similarly the CO data were converted from main beam temperatures to flux densities. Both maps were then scaled to the distance of M\,31 of 780~kpc, requiring changes to the flux densities, pixel and beam sizes. In order to achieve realistic simulations, we additionally changed the source coordinates for Orion to those of M\,31, such that the simulated $(u,v)$ coverage is comparable to that of our actual M\,31 observations.

The resulting images were then used as input for the simulated observations in CASA~5.3, using the tasks simobserve and simanalyze. In these simulations, we assume full tracks with the subcompact configuration of the SMA, to ensure realistic $(u,v)$ coverage. The simulated continuum images of Orion, observed with the SMA {\it as if it was located in M\,31} are shown in Figure~\ref{fig_simori} and compared to the detectable extent of CO emission from this region. The left panel of Figure~\ref{fig_simori} shows the input {\it Planck} 213 GHz continuum emission and the right panel the simulated continuum observations. The blue contours show the depths that would be reached for integration times corresponding to 1, 2 and 4 interferometer tracks, assuming a one track 3 $\sigma$ sensitivity of 0.6 mJy bm$^{-1}$. With one track we are able to detect the Ori A cloud at the 3 $\sigma$ level as an unresolved source. Our simulated one track detection corresponds to a dust mass of 216 $\pm$ 32 M$_\odot$ which for $R_{g2d} = 136$ (appropriate for Orion), corresponds to 2.93 $\times$ 10$^4$ M$_\odot$, roughly 40\% of the total cloud mass reported by Lada et al. (2010) for the Orion A GMC. This corresponds to the total cloud mass known to be above an extinction of A$_V$ $\approx$ 4.0 magnitudes. 

These simulations suggest that continuum sources at our limit of detection in M\,31 are comparable in scale to the local GMCs. 
Moreover, since the dynamic range in dust emission in GMCs is not very high, our simulations suggest the exciting possibility that one would not have to go too much deeper (2- 3 additional tracks) to be able to detect dust emission with the SMA from a very significant fraction, if not the entirety, of a GMC in M31. The simulations also add support to the notion that the CO is detected a lot more easily than the continuum, and that one CO track is sufficient to detect and define the outer boundary of the clouds. 

Finally, we obtain $\alpha_{\rm CO}$ measurements for these simulated observations of the Orion A  cloud in M\,31, using the same procedure that we also applied to our actual observations. For Orion A  we find $\alpha^\prime_{\rm CO-dust} = 0.038\pm0.006$~M$_\odot$ (K km s$^{-1}$ pc$^2$)$^{-1}$, corresponding to $\alpha_{\rm CO}=5.2\pm0.8$~$M_\odot$ (K\,km\,s$^{-1}$pc$^{2}$)$^{-1}$, assuming a gas-to-dust ratio of 136. This is compatible with standard Galactic values and those we find in M31.  Comparison of our simulations with our observations reinforces the notion that the basic physical nature of GMCs in M31 is very similar to that of GMCs in the local region of the Milky Way.

\section{Summary and conclusions}

We have taken advantage of the SMA wideband receiver upgrade to obtain the first continuum detections of the thermal dust emission on sub-GMC scales ($\sim$ 15 pc) within individual GMCs in the Andromeda galaxy. These include the first resolved continuum detections of dust emission from individual GMCs beyond the Magellanic Clouds. 

We detected 23 discrete continuum sources toward 15 target regions. Of these, 19 objects are shown to be sources of thermal dust emission from within individual GMCs; ten of these sources are resolved. Derived dust masses range from approximately 200 -- 1800 M$_\odot$. We also simultaneously obtained high quality in-band measurements of CO (2--1) emission from each region with the same ({\it u,v}) coverage, astrometry and calibration as the continuum observations. We combined these measurements to make direct determinations of the CO conversion factor, $\alpha_{\rm CO}$, in individual GMCs across the disk of M31. Our direct determination of a conversion factor from CO luminosity to dust mass yields $\alpha^\prime_{\rm CO-dust} = 0.042\pm0.018$~$M_\odot$ (K km s$^{-1}$ pc$^2$)$^{-1}$ with a relatively small dispersion. In particular, this value does not appear to be a function of galactocentric radius (at least over a range from 6 -- 12 kpc). Assuming a gas-to-dust ratio of $\sim136$ for comparison with the Milky Way clouds, we find a corresponding $\alpha_{\rm CO} = 5.7 \pm 2.4 M_\odot$~ (K km s$^{-1}$ pc$^2$)$^{-1}$ for the 2-1 transition of CO, which, within the uncertainties, is in excellent agreement with the value (6.1) previously found to characterize Milky Way clouds and the local universe. With this gas-to-dust ratio, the dust emitting regions in these GMCs have total gaseous masses of 24,000 to 240,000 $M_\odot$. While $\alpha^\prime_{\rm CO-dust}$ does not depend on the gas-to-dust ratio, our results were derived assuming a constant dust temperature of 20~K for all the clouds. However, we show that the temperatures can be varied in a reasonable range even within an error budget for $\alpha_{\rm CO}$ of a factor of two. 

Additionally, we demonstrate with simulated observations of an Orion-like cloud in M~31 that there is meaningful parameter overlap between our sample presented here and the local Galactic GMCs. With these results in hand and a growing sample of clouds being observed with the SMA, it will finally be possible place our understanding of the nature of the GMC population of an external galaxy, like M31, on a similar footing to that of the well studied GMCs in the local Milky Way.

\acknowledgments

We thank Chris Faesi for insightful discussions and assistance with verification of CO reduction procedures. 
The Submillimeter Array is a joint project between the Smithsonian Astrophysical Observatory and the Academia Sinica Institute of Astronomy and Astrophysics and is funded by the Smithsonian Institution and the Academia Sinica.

%

\vspace{5mm}
\facilities{Submillimeter Array}


\software{MIRIAD \citep{sau95}, MIR IDL package (https://www.cfa.harvard.edu/rtdc/SMAdata/ process/mir/), CASA \citep{mcm07}}





\vfill
\eject
\bibliography{m31ref}{}
\bibliographystyle{aasjournal}



\end{document}